\documentclass[aps,showpacs,preprint,amsmath,floatfix]{revtex4}

\usepackage{graphicx}
\usepackage{epsfig}
\usepackage{dcolumn}
\usepackage{bm}

\begin{document}

\title{Evidence against manifest right-handed currents in
neutron beta decay}

\author{A.~Garc\'{\i}a}

\thanks{Deceased}

\affiliation{
Departamento de F\'{\i}sica.\\
Centro de Investigaci\'on y de Estudios Avanzados del IPN.\\
A.P. 14-740.\\
M\'exico, D.F., 07000. M\'EXICO.
}

\author{G.~S\'anchez-Col\'on}

\email[]{gsanchez@mda.cinvestav.mx}

\affiliation{
Departamento de F\'{\i}sica Aplicada.\\
Centro de Investigaci\'on y de Estudios Avanzados del IPN.\\
Unidad M\'erida.\\ A.P. 73, Cordemex. \\
M\'erida, Yucat\'an, 97310. M\'EXICO.
}

\date{\today}

\begin{abstract}

Bounds and presence of manifest right handed currents in neutron
beta decay are reviewed. Assuming the unitarity of the
Cabibbo-Kobayashi-Maskawa matrix, current experimental situation
imposes very stringent limits on the mixing angle,
$-0.00077<\zeta<0.00089$, and on the mass eigenstate,
$M_2({\rm GeV})\in(291.4,439.9)$, in contradiction with the established
lower bound on $M_2$.

\end{abstract}


\maketitle

\section{\label{secone}Introduction}

The Standard Model (SM) has its predictive power in
neutron beta decay (n$\beta$d) afflicted by the fact that it has
two free parameters, namely, $V_{\rm ud}$ and $\lambda=g_1/f_1$
(the ratio of the two leading form factors at zero momentum
transfer). In order to make precise predictions, both parameters
should be determined experimentally with great precision. The
observables measured with the best precision in free n$\beta$d
are the transition rate $R$ and the electron-neutron spin
asymmetry $\alpha_e$. In superallowed n$\beta$d $V_{\rm ud}$ can
be determined very precisely. At present, the problem is that
measurements of $\alpha_e$ give two incompatible values. Despite
this difficulty it is still possible to obtain precise
predictions for the region of validity of the SM using the
expressions of the SM for $R$ and $\alpha_e$ (instead of their
experimental values) and the unitarity of the
Cabibbo-Kobayashi-Maskawa (CKM) matrix along with the
experimental values of $V_{\rm us}$ and $V_{\rm ud}$. This
analysis was carried out in Ref.~\cite{garcia08} and the best
prediction of the SM for free n$\beta$d is given in Table~II and
depictured in Fig.~2(a) of this reference.

In this paper we want to extend this approach to study the
bounds and the presence of right handed currents~\cite{beg77}
(RHC) in n$\beta$d. Two new free parameters are introduced, the
mixing angle $\zeta$ of $W_L$ and $W_R$ and the ratio of squares
of the masses of the corresponding mass eigenstates
$\delta=(M_1/M_2)^2$. In addition, we shall use the very precise
current measurement of $V_{\rm ud}$ in nuclear physics, which as
we shall see plays a very important role.

We have assumed that the CKM matrix is common to $W_L$ and $W_R$.
This is referred to as manifest RHC~\cite{beg77}.

\section{\label{sectwo}Expressions and experimental situation}

The SM predicts for the decay rate of n$\beta$d the expression

\begin{equation}
R(10^{-3}\,{\rm s}^{-1}) =
|V_{\rm ud}|^2(0.1897)(1 + 3\lambda^2)(1 + 0.0739 \pm 0.0008)
\label{RSM}
\end{equation}

\noindent
at the level of a precision of $10^{-4}$. The detailed derivation of
Eq.~(\ref{RSM}) is found in Ref.~\cite{garcia01}. The current
experimental value of the neutron mean life~\cite{pdg08} produces
$R_{\rm exp}(10^{-3}\,{\rm s}^{-1})=1.12905(132)$. The
theoretical error in $R$ of 0.0008 is included (recently this
theoretical bias has been
reduced~\cite{czarnecki04,marciano06}). In our analysis this
theoretical error in $R$ is folded into its experimental error
bar, $\sigma_{R}=0.00102$ becomes $\sigma'_{R}=0.00132$ (in
units of $10^{-3}\,{\rm s}^{-1}$). However, it must be stressed
that our analysis is independent of $R_{\rm exp}$ and its error
bar $\sigma'_{R}$ and $\alpha_e$. This is true even though the
neutron mean life is not yet fully converged~\cite{serebrov05}
and the reason for this is that the analysis of
Sec.~\ref{secthree} to obtain the regions of validity of the SM
and the SM with RHC is based on the expressions of $R$ and
$\alpha_e$ instead of their experimental values.

The advantage of the integrated observables $\alpha_{e}$,
$\alpha_{e\nu}$, and $\alpha_{\nu}$, is that their definition
entail only kinematics and do not assume any particular
theoretical approach. The electron neutrino angular correlation
coefficient is defined as $\alpha_{e\nu}=2[
N(\theta_{e\nu}<\pi/2) -
N(\theta_{e\nu}>\pi/2)]/[N(\theta_{e\nu}<\pi/2) +
N(\theta_{e\nu}>\pi/2)]$, where $N(\theta_{e\nu}<\pi/2)$
($N(\theta_{e\nu}>\pi/2)$) is the number of all events with
electron-neutrino pairs emitted in directions that make an angle
between them smaller (greater) than $\pi/2$. Similarly the
electron-neutron spin asymmetry coefficient is defined as
$\alpha_{e}=2[N(\theta_{e}<\pi/2) -
N(\theta_{e}>\pi/2)]/[N(\theta_{e}<\pi/2) +
N(\theta_{e}>\pi/2)]$, where $\theta_{e}$ is the angle between
the electron direction and the polarization direction of the
neutron. Analogous definition is used for the neutrino-neutron
spin asymmetry $\alpha_{\nu}$. Reference~\cite{garcia85} provides
the complete numerically integrated formulas for the decay rate
and angular coefficients.

At the $10^{-4}$ level the SM predicts for the electron-asymmetry
the expression~\cite{garcia06}

\begin{equation}
\alpha_e =
\frac{-0.00021 + 0.2763\lambda - 0.2772\lambda^2}
{0.1897 + 0.5692\lambda^2}.
\label{alphae}
\end{equation}

\noindent
We have chosen a negative sign for $\lambda$ to conform with the
convention of~\cite{pdg08}. The important remark here is that
there is no theoretical uncertainty in $\alpha_e$ at this level
of precision. The reason for this is that the uncertainty
introduced by the model dependence of the contributions of $Z^0$
to the radiative corrections is common to the numerator and
denominator of $\alpha_e$ and cancels away at the $10^{-4}$
level.

The analysis that leads to Eq.~(\ref{alphae}) can be extended to
the neutrino and electron-neutrino asymmetry coefficients,

\begin{equation}
\alpha_{\nu} =
\frac{0.0003 - 0.3794\lambda - 0.2772\lambda^2}
{0.1897 + 0.5692\lambda^2},
\label{alphanu}
\end{equation}

\begin{equation}
\alpha_{e\nu} =
\frac{0.1382 + 0.00054\lambda - 0.1393\lambda^2}
{0.1897 + 0.5692\lambda^2}.
\label{alphaenu}
\end{equation}

It must be stressed that the angular coefficients are free of a
theoretical error at a level of precision of $10^{-4}$. This
accuracy is better than the current experimental precision that
modern experiments allow. The effects of strong interactions,
radiative corrections, and the recoil of the proton have been
included~\cite{garcia06}.

It has remained customary to present experimental results for the
old order zero angular coefficients after all the corrections
contained in $\alpha_e$, $\alpha_{\nu}$, and $\alpha_{e\nu}$,
have been applied to the experimental analysis~\cite{pdg08},

\begin{equation}
A_0 =-
\frac{2\lambda(\lambda+1)}{1+3\lambda^2},
\label{Asub0}
\end{equation}

\begin{equation}
B_0 =
\frac{2\lambda(\lambda-1)}{1+3\lambda^2},
\label{bsub0}
\end{equation}

\begin{equation}
a_0 =
\frac{1-\lambda^2}{1+3\lambda^2}.
\label{asub0}
\end{equation}

\noindent
Also, besides of presenting results for $A_0$ it is
customary to report directly the value for $\lambda$ obtained
from expression~(\ref{Asub0}). Thus, the experimental value of
$\lambda$ is free of theoretical uncertainties at the $10^{-4}$
level. We use this value of $\lambda$ in Eq.~(\ref{alphae}) to
estimate the corresponding value of $\alpha_e$ and its error bar.
By following a similar procedure with eqs.~(\ref{bsub0}) and
(\ref{alphanu}), and (\ref{asub0}) and (\ref{alphaenu}), we
obtain the numerical values of $\alpha_{\nu}$ and
$\alpha_{e\nu}$.

From present experimental results~\cite{pdg08} for the
n$\beta$d order zero angular coefficients, $B_0$, $a_0$, and
$A_0$, the corresponding experimental values of the
integrated angular coefficients are $\alpha^{\rm
exp}_{\nu}=0.9810(30)$, $\alpha^{\rm exp}_{e\nu}=-0.0772(29)$,
and the two conflicting values for $\alpha_e$, $\alpha^{\rm
exp}_e({\rm A})=-0.08809(52)$~\cite{abele02,abele08} and
$\alpha^{\rm exp}_e({\rm
LYB})=-0.08489(65)$~\cite{liaud97,yerozolimsky97,bopp86}.

The expressions of the observables in free n$\beta$d of the SM
including the contributions of RHC with a precision of $10^{-4}$
can be expressed as

\begin{equation}
R=(1.0739) A_{\alpha }
   V_{\text{ud}}^2
   \left(0.1897+(0.5692)
   B_{\alpha }\lambda ^2\right),
\label{R}
\end{equation}

\begin{equation}
\alpha_{e}=\frac{D_{\alpha }
   \left(-0.00021-(0.2763)
   F_{\alpha }\lambda-(0.2772)
   E_{\alpha } \lambda^2\right)}{A_{\alpha }
\left(0.1897+(0.5692)
   B_{\alpha }\lambda^2\right)}
\label{alphaerhc}
\end{equation}

\begin{equation}
\alpha_{\nu}=
\frac{D_{\alpha } \left(0.0003-(0.3794)
     F_{\alpha }\lambda+(0.3795)
   E_{\alpha } \lambda^2\right)}{A_{\alpha }
\left(0.1897+(0.5692)
   B_{\alpha }\lambda^2\right)}
\label{alphanurhc}
\end{equation}

\begin{equation}
\alpha_{e\nu}=
\frac{0.1382+(0.00054)
   C_{\alpha }\lambda-(0.1393)
   B_{\alpha }\lambda^2}{0.1897+(0.5692)
   B_{\alpha }\lambda^2}.
\label{alphaenurhc}
\end{equation}

\noindent
Here~\cite{beg77}, $A_{\alpha}\ldots F_{\alpha}$ contain the
corrections due to RHC. $A_{\alpha}=
2(\eta_{\text{AV}}^2+1)/(\eta_{\text{AA}}^2+2
\eta_{\text{AV}}^2+1)$,
$B_{\alpha}=(\eta_{\text{AA}}^2+\eta_{\text{AV}}^2)
/(\eta_{\text{AV}}^2+1)$, $C_{\alpha}=
(\eta_{\text{AA}}+\eta_{\text{AV}}^2)/(\eta_{\text{AV}}^2+1)$,
$D_{\alpha}=-4\eta_{\text{AV}}/(\eta_{\text{AA}}^2+2
\eta_{\text{AV}}^2+1)$, $E_{\alpha}=\eta_{\text{AA}}$,
$F_{\alpha}=(\eta_{\text{AA}}+1)/2$, where
$\eta_{\text{AA}}=(\delta +\epsilon^2)/(\delta\,\epsilon^2+1)$,
$\eta_{\text{AV}}=-(1-\delta)\epsilon/(\delta\,\epsilon^2+1)$,
with $\epsilon=(1+\tan{\zeta})/(1-\tan{\zeta})$. The numerical
coefficients reamin the same as in
eqs.~(\ref{RSM})--(\ref{alphaenu}).

\section{\label{secthree}Determination of the regions of
validity}

The region where the SM and the SM with RHC (SMR and RHCR,
respectively) remain valid at a $90\%$~CL are determined by forming
a ${\chi}^{2}$ function with the sum of six terms, $((R_{\rm
exp}-R)/\sigma'_{R})^2$, $((\alpha^{\rm
exp}_e-\alpha_e)/\sigma_{\alpha_{e({\rm LYB})}})^2$,
$((\alpha^{\rm
exp}_{\nu}-\alpha_{\nu})/\sigma_{\alpha_{\nu}})^2$,
$((\alpha^{\rm
exp}_{e\nu}-\alpha_{e\nu})/\sigma_{\alpha_{e\nu}})^2$, $((V^{\rm
exp}_{\rm us}-V_{\rm us}\,A^{1/2}_\alpha)/\sigma_{V_{\rm
us}})^2$, and $((V^{\rm exp}_{\rm ub}-V_{\rm
ub}\,A^{1/2}_\alpha)/\sigma_{V_{\rm ub}})^2$, where $V_{\rm ub}
= \sqrt{1 - V^2_{\rm ud} - V^2_{\rm us}}$, and then minimizing
the ${\chi}^{2}$ at a lattice of points $(\alpha^{\rm exp}_e,
R_{\rm exp})$ within a rectangle that covers $\pm 3\sigma'_{R}$
around $R_{\rm exp}$ and a range for $\alpha^{\rm exp}_e$
covering $\alpha^{\rm exp}_e({\rm A})$ and $\alpha^{\rm
exp}_e({\rm LYB})$. The values of $\sigma'_{R}$ and
$\sigma_{\alpha_{e({\rm LYB})}}$ can also be reduced from their
currents values of $0.00132\times 10^{-3}\,{\rm s}^{-1}$ and
0.00065 to one-tenth of these values which run into the
theoretical error bars of $10^{-4}$. The free parameters varied
at each $(\alpha^{\rm exp}_e, R_{\rm exp})$ point are $\lambda$,
$V_{\rm ud}$, and $V_{\rm us}$ for the SMR and $\lambda$, $V_{\rm
ud}$, $V_{\rm us}$, $\zeta$, and $\delta$ for the RHCR. In
addition, we shall add a seventh constraint $((V^{\rm exp}_{\rm
ud}({\rm NP})-V_{\rm ud}\,A^{1/2}_\alpha)/\sigma_{V_{\rm
ud}})^2$ to ${\chi}^{2}$ which incorporates the experimental
nuclear physics (NP) value of $V^{\rm exp}_{\rm ud}({\rm
NP})=0.97418(27)$.

\begin{figure}
\centerline{\psfig{file=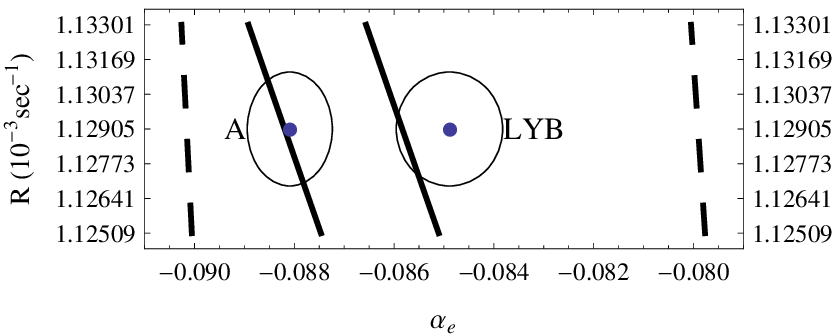,width=6.4in}}
\caption{\label{fig_1}
$90\%$~CL SMR (solid lines) and RHCR (dashed lines) without
the $V^{\rm NP}_{\rm ud}$ constraint included. The $90\%$~CL region
around the current central values of $\alpha^{\rm exp}_e$
and $R_{\rm exp}$ are also displayed.
}
\end{figure}

\begin{figure}
\centerline{\psfig{file=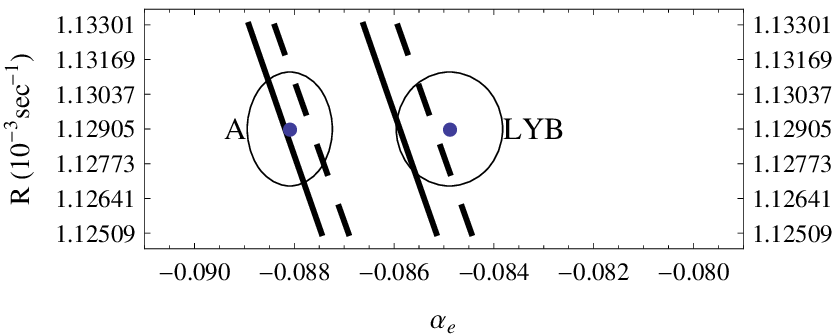,width=6.4in}}
\caption{\label{fig_2}
$90\%$~CL SMR (solid lines) and RHCR (dashed lines) with
the $V^{\rm NP}_{\rm ud}$ constraint included. The $90\%$~CL region
around the current central values of $\alpha^{\rm exp}_e$
and $R_{\rm exp}$ are also displayed.
}
\end{figure}

The numerical results are displayed in Table~\ref{tablaIII2}
without the $V^{\rm NP}_{\rm ud}$ constraint and in
Table~\ref{tablaIV2p} with the $V^{\rm NP}_{\rm ud}$ constraint
included. The corresponding $90\%$~CL SMR and RHCR are depicted in
Figs.~\ref{fig_1} and \ref{fig_2}.

\section{\label{secfour}Discussion}

In Table~\ref{tablaIII2} the constraint of $V^{\rm NP}_{\rm ud}$
is not enforced, while in Table~\ref{tablaIV2p} this constraint
is operative. In both tables in each entry the upper numbers obey
the constraint of $\alpha^{\rm exp}_{\nu}$ and the lower ones do
not obey it. The last two columns
give the $90\%$~CL bounds on the two free parameters of manifest RHC.

The $\chi^2$ of the SM predictions in both tables show a
discrepancy of 2.2 standard deviation. One can see that such a
discrepancy is saturated by $\chi^2(\alpha_{\nu})$. The presence
or absence of the $V^{\rm NP}_{\rm ud}$ constraint plays no role
in this discrepancy. When RHC are allowed in, one can appreciate
the relevance of $V^{\rm NP}_{\rm ud}$. The bounds on $\zeta$ are
reduced and made very uniform when $V^{\rm NP}_{\rm ud}$
constrains $\chi^2$. The ranges for $\zeta$ in
Table~\ref{tablaIII2} are negative at the top five entries and
only in the last two at the bottom $\zeta=0$ is allowed. The
length $\Delta\zeta$ of these ranges is around 0.00660. In
contrast, in Table~\ref{tablaIV2p} the ranges for $\zeta$ are
quite symmetric around $\zeta=0$ and have $\Delta\zeta$ of
0.00166, approximately one-fourth of the length when $V^{\rm
NP}_{\rm ud}$ is not operative. One can also see in the lower
numbers that whether $\alpha^{\rm exp}_{\nu}$ is enforced or not
makes no difference. One can, then, conclude that the bounds of
$\zeta$, typically of

\begin{equation}
\zeta\in(-0.00077,0.00089)
\label{zeta}
\end{equation}

\noindent
are imposed solely by $V^{\rm NP}_{\rm ud}$, $V_{\rm us}$, and
the unitarity of the CKM matrix. These bounds may be compared
with previous ones. In Ref.~\cite{aquino91} one had
$\zeta\in(-0.00060,0.00280)$ with a $\Delta\zeta=0.0340$. The
range~(\ref{zeta}) is more symmetric and has half the length.

The bounds on $\delta$ are practically independent of $V^{\rm
NP}_{\rm ud}$, but they are very dependent on $\alpha^{\rm
exp}_{\nu}$ as can be seen by comparing the upper and the lower
numbers. Actually, the upper bound on $\delta$, at around 0.076
is also almost independent of $\alpha^{\rm exp}_{\nu}$. It is the
lower bound on $\delta$ that is very sensitive on $\alpha^{\rm
exp}_{\nu}$. In Table~\ref{tablaIV2p} it varies from about
$\delta\approx 0.033>0$ to about $\delta\approx -0.1020<0$,
according to whether $\alpha^{\rm exp}_{\nu}$ is operative or
not. Of course, a negative $\delta$ is meaningless and the actual
lower bound should be $\delta=0$, which makes the range for
$\delta$ an upper bound only. One can conclude that $\alpha^{\rm
exp}_{\nu}$ imposes the $90\%$~CL range of

\begin{equation}
\delta\in(0.0334,0.0761)
\label{delta}
\end{equation}

\noindent
upon $\delta$.

At this point one should translate~(\ref{delta}) into a range
for $M_2$. One has

\begin{equation}
M_2({\rm GeV})\in(291.4,439.9).
\label{m2}
\end{equation}

\noindent
Range~(\ref{m2}) shows vividly how effective is $\alpha_{\nu}$
for setting an upper bound on $M_2$. It also means that manifest
RHC are detected in n$\beta$d. However, one already knows that
lower bounds on $M_2$ have been established. At present one may
accept as a conservative lower bound $M_2>715$~GeV~\cite{pdg08}.
This is in clear contradiction with range~(\ref{m2}).

In order to better understand this situation we have prepared
another table, Table~\ref{tablaV}. We are interested in
appreciating what refined measurements of $\alpha^{\rm
exp}_{\nu}$ may produce in, hopefully, the near future. We assume
that the error bar $\sigma_{\alpha_{\nu}}$ is reduced to
one-tenth of its current value. That is, we assume
$\sigma_{\alpha_{\nu}}=0.00030$ and we vary the central value
$\alpha^{\rm exp}_{\nu}$ from 0.98100 to 0.98760. We keep $R_{\rm
exp}$, $V^{\rm exp}_{\rm ud}({\rm NP})$, and $V^{\rm exp}_{\rm
us}$ at their current central values and error bars. The results
are displayed in Table~\ref{tablaV}, in steps of 0.00060.

As can be seen in the last column of Table~\ref{tablaV}, at the
$90\%$~CL only when the experimental value of $\alpha_{\nu}$ is
greater than 0.9870 the upper bound obtained for $M_2$ is not
ruled out by its present established lower bound. For
$\alpha^{\rm exp}_{\nu}\ge 0.9876$ the central value for $\delta$ is
compatible with zero. One can conclude that a clean signal of
manifest RHC can be obtained only if future measurements of
$\alpha^{\rm exp}_{\nu}$ find it in the range

\begin{equation}
\alpha^{\rm exp}_{\nu}\in(0.9870,0.9876).
\label{alfanu}
\end{equation}

\section{\label{secfive}Conclusions}

The current experimental situation in n$\beta$d and in the lower
bounds on $M_2$ lead one to conclude that manifest RHC run into a
contradiction, that leads one to conclude that manifest RHC are
strongly eliminated as a possibility of physics beyond the SM. The
experimental quantity which leads to this conclusion is the
current value of $\alpha_{\nu}$.

However, future refined experiments may correct the current
situation provided two conditions are met: (1) $\alpha_{\nu}$ is
found within range~(\ref{alfanu}) and (2) $\alpha_{e}$ is found
in the future in the range $\alpha^{\rm
exp}_{e}\in(-0.08570,-0.08717)$ of Table~\ref{tablaIV2p}. If
either of these conditions fail, then manifest RHC will be
strongly eliminated. Of course, other forms of new physics could
be detected by $\alpha_{\nu}$, as can be appreciated by the
values of $\chi^2$ in the SM case in Table~\ref{tablaV}.

As a final remark, it is not idle to emphasize the importance of
refined very precise measurements of the observables in n$\beta$d.

\begin{acknowledgments}

The authors would like to thank CONACyT (M\'exico) for partial
support.

\end{acknowledgments}

\clearpage

{\squeezetable

\begin{table}

\caption{The minimum of $\chi^2$, its corresponding value of
$\alpha_e$, the prediction for $\alpha_{\nu}$,
and the partial contribution from $\alpha_{\nu}$
to $\chi^2$ for seven values of $R_{\rm exp}$ (in units of
$10^{-3}{\rm sec}^{-1}$) without the $V^{\rm NP}_{\rm ud}$
constraint. The upper numbers obey the constraint of
$\alpha^{\rm exp}_{\nu}$ and the lower ones do not obey it. The
last two columns give the $90\%$~CL bounds on the two free parameters
of manifest RHC, $\zeta$ and $\delta$, respectively.
\label{tablaIII2}}

\begin{ruledtabular}

\begin{tabular}{c|cccc|cccccc}

 & \multicolumn{4}{c|}{SM}  & \multicolumn{6}{c}{RHC} \\

 & value & & prediction & & value & & prediction & & & \\

$R$ & $\alpha_e$ & $\chi^2$ & $\alpha_{\nu}$ &
$\chi^2(\alpha_{\nu})$ & $\alpha_e$ & $\chi^2$ & $\alpha_{\nu}$
& $\chi^2(\alpha_{\nu})$ & $\zeta$ & $\delta=(M_1/M_2)^2$ \\

\hline

1.13301 & $-$0.08772 & 5.32 & 0.98759 & 4.82 & $-$0.08497 &
$10^{-5}$ & 0.98100 & $10^{-7}$ & $(-0.00924,-0.00263)$ &
$(0.0384,0.0812)$ \\

 & $-$0.08772 & 0.50 &  &  & $-$0.08497 & $10^{-6}$ &  &  &
$(-0.00953,0.00884)$ & $(-0.2560,0.1662)$ \\

1.13169 & $-$0.08752 & 5.33 & 0.98765 & 4.92 & $-$0.08497 &
$10^{-4}$ & 0.98100 & $10^{-7}$ & $(-0.00856,-0.00195)$ &
$(0.0380,0.0808)$ \\

 & $-$0.08749 & 0.41 &  &  & $-$0.08497 & $10^{-5}$ &  &  &
$(-0.00890,0.00953)$ & $(-0.2624,0.1645)$ \\

1.13037 & $-$0.08726 & 5.35 & 0.98772 & 5.02 & $-$0.08492 &
$10^{-5}$ & 0.98100 & $10^{-7}$ & $(-0.00801,-0.00140)$ &
$(0.0378,0.0804)$ \\

 & $-$0.08723 & 0.33 &  &  & $-$0.08492 & $10^{-7}$ &  &  &
$(-0.00839,0.01022)$ & $(-0.2701,0.1636)$ \\

1.12905 & $-$0.08700 & 5.38 & 0.98779 & 5.12 & $-$0.08487 &
$10^{-5}$ & 0.98100 & $10^{-7}$ & $(-0.00746,-0.00084)$ &
$(0.0377,0.0802)$ \\

 & $-$0.08700 & 0.25 &  &  & $-$0.08487 & $10^{-6}$ &  &  &
$(-0.00788,0.01095)$ & $(-0.2782,0.1626)$ \\

1.12773 & $-$0.08679 & 5.42 & 0.98786 & 5.23 & $-$0.08483 &
$10^{-5}$ & 0.98100 & $10^{-8}$ & $(-0.00691,-0.00029)$ &
$(0.0375,0.0799)$ \\

 & $-$0.08676 & 0.19 &  &  & $-$0.08483 & $10^{-6}$ &  &  &
$(-0.00738,0.01168)$ & $(-0.2867,0.1617)$ \\

1.12641 & $-$0.08653 & 5.47 & 0.98793 & 5.33 & $-$0.08479 &
$10^{-5}$ & 0.98100 & $10^{-7}$ & $(-0.00633,0.00029)$ &
$(0.0372,0.0796)$ \\

 & $-$0.08650 & 0.14 &  &  & $-$0.08479 & $10^{-5}$ &  &  &
$(-0.00686,0.01251)$ & $(-0.2956,0.1607)$ \\

1.12509 & $-$0.08627 & 5.53 & 0.98799 & 5.44 & $-$0.08473 &
$10^{-6}$ & 0.98100 & $10^{-9}$ & $(-0.00581,0.00082)$ &
$(0.0371,0.0794)$ \\

 & $-$0.08627 & 0.09 &  &  & $-$0.08473 & $10^{-6}$ &  &  &
$(-0.00638,0.01329)$ & $(-0.3055,0.1599)$ \\

\end{tabular}

\end{ruledtabular}

\end{table}}

\clearpage

{\squeezetable

\begin{table}

\caption{The minimum of $\chi^2$, its corresponding value of
$\alpha_e$, the prediction for $\alpha_{\nu}$,
and the partial contribution from $\alpha_{\nu}$
to $\chi^2$ for seven values of $R_{\rm exp}$ (in units of
$10^{-3}{\rm sec}^{-1}$) with the $V^{\rm NP}_{\rm ud}$
constraint. The upper numbers obey the constraint of
$\alpha^{\rm exp}_{\nu}$ and the lower ones do not obey it. The
last two columns give the $90\%$~CL bounds on the two free parameters
of manifest RHC, $\zeta$ and $\delta$, respectively.
\label{tablaIV2p}}

\begin{ruledtabular}

\begin{tabular}{c|cccc|cccccc}

 & \multicolumn{4}{c|}{SM}  & \multicolumn{6}{c}{RHC} \\

 & value & & prediction & & value & & prediction & & & \\

$R$ & $\alpha _e$ & $\chi^2$ & $\alpha
_{\nu}$ & $\chi^2(\alpha _{\nu})$ & $\alpha _e$ & $\chi^2$ &
$\alpha_{\nu}$ & $\chi^2(\alpha _{\nu})$ & $\zeta$ &
$\delta=(M_1/M_2)^2$ \\

\hline

1.13301 & $-$0.08778 & 5.34 & 0.98758 & 4.81 & $-$0.08717 & 0.51 &
0.98100 & $10^{-8}$ & $(-0.000785,0.000890)$ & $(0.0318,0.0754)$ \\

 & $-$0.08775 & 0.52 &  &  & $-$0.08717 & 0.51 & & &
$(-0.000769,0.000892)$ & $(-0.1011,0.0997)$ \\

1.13169 & $-$0.08752 & 5.35 & 0.98765 & 4.91 & $-$0.08694 & 0.42 &
0.98104 & $10^{-4}$ & $(-0.000784,0.000891)$ & $(0.0320,0.0755)$ \\

 & $-$0.08752 & 0.43 &  &  & $-$0.08694 & 0.42 & & &
$(-0.000768,0.000892)$ & $(-0.1007,0.0992)$ \\

1.13037 & $-$0.08726 & 5.36 & 0.98772 & 5.01 & $-$0.08668 & 0.34 &
0.98102 & $10^{-5}$ & $(-0.000784,0.000891)$ & $(0.0327,0.0758)$ \\

 & $-$0.08726 & 0.35 &  &  & $-$0.08668 & 0.34 & & &
$(-0.000784,0.000893)$ & $(-0.1014,0.1000)$ \\

1.12905 & $-$0.08705 & 5.40 & 0.98778 & 5.11 & $-$0.08642 & 0.26 &
0.98100 & $10^{-7}$ & $(-0.000768,0.000891)$ & $(0.0334,0.0761)$ \\

 & $-$0.08702 & 0.27 &  &  & $-$0.08642 & 0.26 & & &
$(-0.000768,0.000894)$ & $(-0.1020,0.1005)$ \\

1.12773 & $-$0.08679 & 5.44 & 0.98785 & 5.22 & $-$0.08619 & 0.20 &
0.98102 & $10^{-5}$ & $(-0.000767,0.000892)$ & $(0.0337,0.0762)$ \\

 & $-$0.08679 & 0.21 &  &  & $-$0.08619 & 0.20 & & &
$(-0.000767,0.000894)$ & $(-0.1019,0.1004)$ \\

1.12641 & $-$0.08653 & 5.49 & 0.98792 & 5.32 & $-$0.08596 & 0.15 &
0.98104 & $10^{-4}$ & $(-0.000766,0.000893)$ & $(0.0341,0.0764)$ \\

 & $-$0.08653 & 0.15 &  &  & $-$0.08596 & 0.14 & & &
$(-0.000766,0.000895)$ & $(-0.1018,0.1003)$ \\

1.12509 & $-$0.08632 & 5.55 & 0.98799 & 5.43 & $-$0.08570 & 0.10 &
0.98102 & $10^{-5}$ & $(-0.000766,0.000894)$ & $(0.0347,0.0767)$ \\

 & $-$0.08629 & 0.11 &  &  & $-$0.08570 & 0.10 & & &
$(-0.000766,0.000895)$ & $(-0.1024,0.1009)$ \\

\end{tabular}

\end{ruledtabular}

\end{table}}

\clearpage

{\turnpage

\begin{table}

\caption{The minimum of $\chi^2$, its corresponding value of
$\alpha_e$, the prediction for $\alpha_{\nu}$,
and the partial contribution from $\alpha_{\nu}$
to $\chi^2$ for several values of $\alpha^{\rm exp}_{\nu}$ with
the error bar $\sigma_{\alpha_{\nu}}$ reduced to one-tenth of its
current value. $R_{\rm exp}$, $V^{\rm exp}_{\rm ud}({\rm NP})$,
and $V^{\rm exp}_{\rm us}$ are kept at their current central
values and error bars. The last three columns give the
$90\%$~CL bounds on the two free parameters of manifest RHC,
$\zeta$ and $\delta$, and the corresponding bounds on $M_2$,
respectively. \label{tablaV}}

\begin{ruledtabular}

\begin{tabular}{c|cccc|ccccccc}

 & \multicolumn{4}{c|}{SM}  & \multicolumn{7}{c}{RHC} \\

 & value & & prediction & & value & & prediction & & & & \\

$\alpha _{\nu}$ & $\alpha _e$ & $\chi^2$ & $\alpha _{\nu}$ &
$\chi^2(\alpha _{\nu})$ & $\alpha _e$ & $\chi^2$ &
$\alpha_{\nu}$ & $\chi^2(\alpha _{\nu})$ & $\zeta$ &
$\delta=(M_1/M_2)^2$ & $M_2\,({\rm GeV})$ \\

\hline

0.9810 & $-$0.08840 & 482.99 & 0.98740 & 454.85 &
$-$0.08642 & 0.26 & 0.9810 & $10^{-9}$ & $(-0.000768,0.000891)$ &
$(0.0564, 0.0609)$ &
$(325.8,338.5)$\\

0.9816 & $-$0.08827 & 401.45 & 0.98743 & 378.08 &
$-$0.08648 & 0.26 & 0.9816 & $10^{-8}$ & $(-0.000767,0.000891)$ &
$(0.0536, 0.0583)$ &
$(332.9,347.2)$\\

0.9822 & $-$0.08817 & 327.38 & 0.98747 & 308.23 &
$-$0.08653 & 0.26 & 0.9822 & $10^{-8}$ & $(-0.000767,0.000890)$ &
$(0.0507, 0.0556)$ &
$(340.9,357.0)$\\

0.9828 & $-$0.08804 & 260.98 & 0.98750 & 245.69 &
$-$0.08658 & 0.26 & 0.9828 & $10^{-7}$ & $(-0.000766,0.000889)$ &
$(0.0476, 0.0528)$ &
$(349.8,368.5)$\\

0.9834 & $-$0.08791 & 202.05 & 0.98754 & 190.19 &
$-$0.08663 & 0.26 & 0.9834 & $10^{-7}$ & $(-0.000766,0.000889)$ &
$(0.0443, 0.0498)$ &
$(360.2,381.9)$\\

0.9840 & $-$0.08780 & 150.65 & 0.98757 & 141.71 &
$-$0.08668 & 0.26 & 0.9840 & $10^{-7}$ & $(-0.000765,0.000888)$ &
$(0.0407, 0.0467)$ &
$(372.0,398.5)$\\

0.9846 & $-$0.08767 & 106.80 & 0.98761 & 100.40 &
$-$0.08674 & 0.26 & 0.9846 & $10^{-7}$ & $(-0.000765,0.000887)$ &
$(0.0368, 0.0433)$ &
$(386.3,419.1)$\\

0.9852 & $-$0.08754 & 70.49 & 0.98764 & 66.20 &
$-$0.08679 & 0.26 & 0.9852 & $10^{-7}$ & $(-0.000764,0.000887)$ &
$(0.0324, 0.0396)$ &
$(404.0,446.6)$\\

0.9858 & $-$0.08741 & 41.72 & 0.98768 & 39.09 &
$-$0.08684 & 0.26 & 0.9858 & $10^{-7}$ & $(-0.000764,0.000886)$ &
$(0.0273, 0.0355)$ &
$(426.7,486.5)$\\

0.9864 & $-$0.08731 & 20.49 & 0.98771 & 19.05 &
$-$0.08689 & 0.26 & 0.9864 & $10^{-7}$ & $(-0.000763,0.000885)$ &
$(0.0211, 0.0310)$ &
$(456.6,553.4)$\\

0.9870 & $-$0.08718 & 6.80 & 0.98774 & 6.15 &
$-$0.08694 & 0.26 & 0.9870 & $10^{-7}$ & $(-0.000762,0.000868)$ &
$(0.0119, 0.0256)$ &
$(502.4,736.9)$\\

0.9876 & $-$0.08705 & 0.65 & 0.98778 & 0.35 &
$-$0.08702 & 0.26 & 0.9876 & $10^{-7}$ & $(-0.000758,0.000884)$ &
$(-0.0187, 0.0187)$ &
$>587.9$\\
 
\end{tabular}

\end{ruledtabular}

\end{table}}

\clearpage

\end{document}